\documentclass[%
 reprint,longbibliography,preprintnumbers,
nofootinbib,
 amsmath,amssymb,
 aps
]{revtex4-1}
\pdfoutput=1
\usepackage{graphicx}
\usepackage[utf8]{inputenc}
\usepackage{flushend}
\usepackage{dcolumn}
\usepackage{bm}
\usepackage{balance}

\usepackage[normalem]{ulem}

\usepackage[colorlinks = true,
            linkcolor = blue,
            urlcolor  = blue,
            citecolor =green,
            anchorcolor = blue]{hyperref}
\usepackage{verbatim}
\usepackage{color,ulem}
\usepackage[english]{babel}
\usepackage{lipsum} 

\usepackage[utf8]{inputenc}
\input Starburst.fd
\newcommand*\initfamily{\usefont{U}{Starburst}{xl}{n}}\initfamily

\newcommand{\beq}{\begin{eqnarray}}
\newcommand{\eeq}{\end{eqnarray}}
\usepackage{amsmath}
\usepackage{tikz}
\usetikzlibrary{decorations.pathmorphing}
\usetikzlibrary{shapes.misc}
\tikzset{cross/.style={cross out, draw=black, minimum size=8*(#1-\pgflinewidth), inner sep=0pt, outer sep=0pt},
cross/.default={1pt}}
\usetikzlibrary{patterns,math}
\begin{document}

\title{\huge  Theory of Superconductivity Mediated by Topological Phonons}

\author{\textbf{Daniele Di Miceli}$^{1}$}%
 \email{daniele.dimiceli@studenti.unimi.it}
 \author{\textbf{Chandan Setty}$^{2}$}%
 \email{csetty@rice.edu}
\author{\textbf{Alessio Zaccone}$^{1,3}$}%
 \email{alessio.zaccone@unimi.it}

 \vspace{1cm}
 
\affiliation{$^{1}$
Department of Physics ``A. Pontremoli'', University of Milan, via Celoria 16, 20133 Milan, Italy}
\affiliation{$^{2}$
Department of Physics and Astronomy, Rice Center for Quantum Materials, Rice University, Houston, Texas 77005, USA}
\affiliation{$^{3}$
Cavendish Laboratory, University of Cambridge, JJ Thomson Avenue, CB30HE Cambridge, United Kingdom}

\begin{abstract}
    Topological phononic insulators are the counterpart of three-dimensional quantum spin Hall insulators in phononic systems and, as such, their topological surfaces are characterized by Dirac cone-shaped gapless edge states arising as a consequence of a bulk-boundary correspondence. We propose a theoretical framework for the possible superconducting phase in these materials, where the attractive interaction between electrons is mediated by topological phonons in nontrivial boundary modes.
    Within the BCS limit, we develop a self-consistent two-band gap equation, whose solutions show that the superconducting critical temperature has a non-monotonic behaviour with respect to the phononic frequency in the Kramers-like point. This remarkable behaviour is produced by a resonance, that occurs when electrons and phonons on the topological surfaces have the same energy: this effectively increases the electron-phonon interaction and hence the Cooper pair binding energy, thus establishing an optimal condition for the superconducting phase. With this mechanism, the $T_{c}$ can be increased by well over a factor two, and the maximum enhancement occurs in the degenerate phononic flat-band limit.
\end{abstract}

\maketitle


The discovery of a topological classification for electronic band systems \cite{Kane_review, TIs_review} has shed new light also on the topological phononic systems, leading to the extension of the topological framework over phononic states \cite{topological_phononics}. 
Similar to electrons, phononic quantum Hall-like states can be hosted in time-reversal symmetry breaking phases \cite{phonon_diode, phonon_Hall_effect}, while quantum spin Hall-like ones require novel degrees of freedom to reproduce the Kramers doublet at the time-reversal invariant momenta \cite{acoustic_TIs, acoustic_topological_phases, Acoustic_analogue_TIs}.
Specifically, in three-dimensional lattices, the introduction of a crystalline-protected pseudospin degree of freedom provides the required energy degeneracy \cite{crystalline_TIs, realization_TCIs}, leading to a time-reversal invariant topological classification for phonons.
Nontrivial states thus obtained are referred to as ``phononic topological insulators'' \cite{phononic_TIs}, being the counterpart of 3D topological insulators in phononic materials \cite{TIs_in_3D, Topological_Insulators_Superconductors}. Similarly to electrons \cite{Dirac-cone_states, Dirac-cone_states2}, their topological surfaces are characterized by Dirac cone-shaped gapless \emph{edge states}, which arise as a consequence of a bulk-boundary correspondence \cite{Kane_review} and have great research interest because of their unconventional transport properties \cite{phononic_TIs}.

The purpose of this Letter is to investigate the role of phonons in topological boundary states as mediators of the superconducting interaction between electrons. We propose a theoretical framework for the possible superconducting phase where the Cooper pairing is mediated by the topological phononic edge states. We describe the nontrivial electron pairing through a self-consistent gap equation within the BCS limit \cite{BCS}, where the phonon dynamics is accounted for by a suitable propagator \cite{Mahan, Fetter, Marsiglio-Carbotte}. 
Numerical solutions to this gap equation show that the critical superconducting temperature $T_c$ displays a \emph{non-monotonic} behaviour as a function of the frequency parameter $\omega_0$, namely the phonon frequency at the Kramers-like point. Moreover, we find that the shape of the topological phononic modes affects the superconducting pairing in such a way that the highest peak of the critical temperature is observed in correspondence of flat degenerate bands.
The optimal $\omega_0$, corresponding to the maximum $T_c$ enhancement in the surface of the material, is related to a resonance effect that occurs when the (standard) electrons and the topological phonon states at the interface have the \emph{same} energy. This effectively increases the coupling constant and hence the Cooper pair binding energy, thus establishing an optimal condition for the superconducting phase. 
In agreement with our numerical results, the overall effect of resonance decreases with increasing the slope of the Dirac cone-shaped bands, implying that the highest critical temperature is produced by flat-band degenerate modes.

\paragraph*{Theoretical Model}

\begin{figure}
    \centering
    \includegraphics[width=\linewidth]{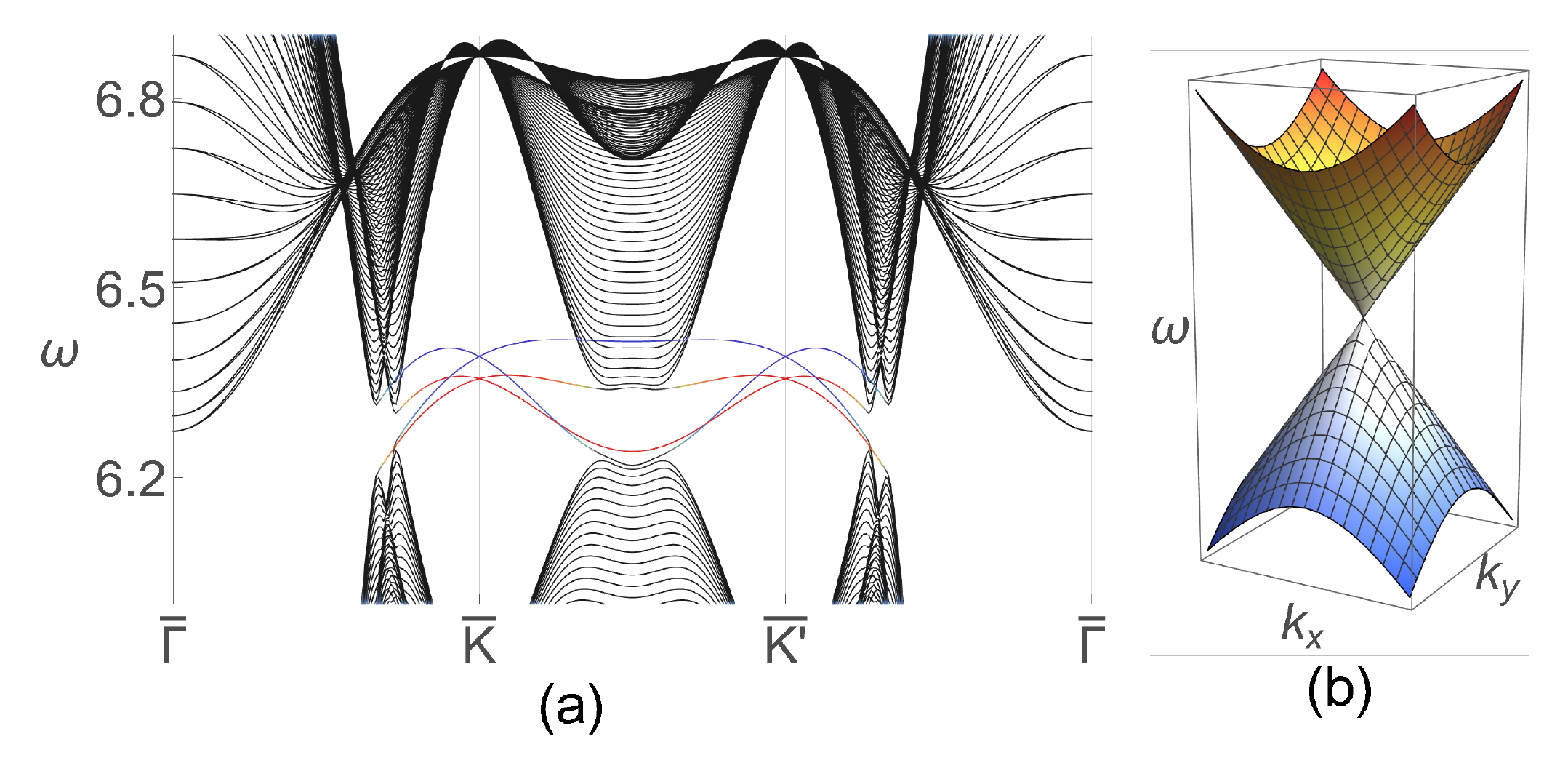}
    \caption{
    (a) Surface and bulk vibrational modes for a phononic topological insulator, computed through the tight-binding scheme proposed in \cite{phononic_TIs}. Black lines represent bulk modes, while red and blue lines stand for topological phononic states that live in the upper and lower surfaces of a finite crystal, respectively.
    (b) Schematic Dirac-cone shaped edge states described by Eq.\eqref{eq:Hamiltonian} as an approximation to describe crossing bands near the high symmetry points $\overline{\mathbf{K}}$ and $\overline{\mathbf{K}}'$. The vertices of the upper and lower cones coincide at the degenerate Kramers-like point with frequency $\omega_0$. The slope of the crossing bands is given by the Dirac velocity $v_D$.
    }
    \label{fig:phonon dispersion}
\end{figure}

Topological insulating states for phonons were theoretically predicted within a 3D triangular crystal \cite{phononic_TIs}, where the in-plane lattice vibrations were investigated through a tight-binding scheme \cite{topological_phononics, phonon_diode}, allowing one to span a wide range of coupling parameters and find suitable conditions for a topological phase transition.
Phononic modes computed for a finite system in a topological insulating-like state are depicted in Fig.\ref{fig:phonon dispersion} (a) as a function of a path joining the high-symmetry points in the Brillouin zone.
Black lines represent bulk bands, while red and blue colours stand for vibrational topological states confined in the upper and lower surfaces of the crystal, respectively.
A single pair of gapless Dirac cone-shaped edge bands \cite{Acoustic_analogue_TIs, phononic_TIs} is located near the high-symmetry points $\overline{\mathbf{K}}$ and $\overline{\mathbf{K}}'$, where the Kramers-like degeneracy is guaranteed by the lattice rotational symmetry \cite{crystalline_TIs}.
Near these special momenta, the surface bands have the structure schematically depicted in Fig.\ref{fig:phonon dispersion} (b), and they can be effectively described by a Dirac-like Hamiltonian, which, in terms of the degenerate momentum, takes the form \cite{phononic_TIs}
\begin{equation}\label{eq:Hamiltonian}
    \mathcal{H}_{surf} = \omega_0 \sigma_0 + v_D (k_x \sigma_x + k_y \sigma_y ) \,,
\end{equation}
where $\sigma_i$ is a complete set of Pauli matrices, and the band parameters $\omega_0$ and $v_D$ are phonon frequency and group velocity at the high-symmetry Kramers-like point, respectively. 
The parameter $\omega_0$ is the ``frequency shift'', which quantifies the energy at the Kramers point where upper and lower bands cross. The parameter $v_D$ represents the slope of the Dirac cone-shaped bands, which become flat degenerate modes in the limiting case where $v_D=0$.
In Eq.\eqref{eq:Hamiltonian}, the appearance of a mass term $M \sigma_z$ introduces a full gap into the phonon spectrum, characterizing a phase transition from topological to ordinary state \cite{Kane_review}. We thus selected $M=0$ to describe nontrivial gapless modes. 

The superconducting many-body interaction between electrons and such boundary topological phonon states can be accounted for in the form of a frequency-dependent electron \emph{self-energy} \cite{Mahan,Fetter}, giving rise to the retardation effects that have a great impact on the macroscopic electron motion \cite{Eliashberg_review, Marsiglio-Carbotte}. 
The electron self-energy can be expressed through the Migdal-Eliashberg diagrammatic approximation \cite{Eliashberg_review, Marsiglio-Carbotte, E-ph_Coupling, Eliashberg_Coupling_Phonons, Migdal}, where all the first order processes are taken into account.
Within the BCS \cite{BCS} weak-coupling limit, the electron self-energy reduces to the simpler gap function $\Delta(\mathbf{k}, i\omega_n)$, thus we can easily derive the following self-consistent gap equation \cite{Eliashberg_renormalization, Enhanced_superconductivity}
\begin{equation}\label{eq:starting_gap_eq}
\begin{gathered}
    \Delta(\mathbf{k},i\omega_n)=-\frac{1}{S\beta} \sum_{\mathbf{k'},m} g_\lambda D_{\lambda\lambda'}(\mathbf{k-k'}, i\omega_n-i\omega_m) g_{\lambda'} \\
    \times \frac{\Delta(\mathbf{k'},i\omega_m)}{\omega_m^2+\xi_\mathbf{k'}^2+\Delta(\mathbf{k'},i\omega_m)^2} \,.    
\end{gathered}
\end{equation}
Here, $S$ is the surface of the topological interfaces, $\beta$ is the inverse temperature, $g_\lambda$ is a constant attractive interaction between electrons and phonons and $D_{\lambda \lambda'}(\mathbf{k},i\nu)$ is the matrix phonon propagator. 
The frequency sum extends over the Matsubara fermionic frequencies $\omega_n = (2n+1) \pi/\beta$, while the momentum one runs over the bi-dimensional Brillouin zone boundary describing the interface states of a phononic topological insulator.
Repeated indices $\lambda$ and $\lambda'$, referred to distinct vibrational states, are summed according to the Einstein convention.
For the sake of simplicity, we assume a constant and frequency-independent gap function $\Delta(\mathbf{k}, i\omega_n) = \Delta$, that allows us to cancel the order parameter in the numerator on both sides of Eq.\eqref{eq:starting_gap_eq} and eliminate the $\omega_n$ dependence, to yield
\begin{equation}
    1=-\frac{1}{S\beta} \sum_{\mathbf{k},m} g_\lambda D_{\lambda\lambda'}(\mathbf{k}, -i\nu_m) g_{\lambda'}
    \frac{1}{\omega_m^2+\xi_\mathbf{k}^2+\Delta^2} \,,
\end{equation}
where $\nu_n = 2n \pi/\beta$ is the bosonic Matsubara frequency.
Here $\xi_\mathbf{k}$ is the free-electron dispersion, which we choose to be quadratic with a chemical potential $\mu$ in order to describe the superconducting coupling between metallic electrons and multi-band phonons.

The matrix Green function $D_{\lambda \lambda'}(\mathbf{k},i\omega)$, that takes into account the whole phonon dynamics in the topological interfaces, can be computed through 
\begin{equation}
    D(\mathbf{k}, \omega) = \sum_j \frac{u_{\mathbf{k}j} u^\dagger_{\mathbf{k}j}}
    {\omega-\epsilon_{\mathbf{k}j}} \,,
\end{equation}
where the sum extends over all the eigenstates $u_{\mathbf{k} j}$ and eigenenergies $\epsilon_{\mathbf{k} j}$ of the surface Hamiltonian in Eq.\eqref{eq:Hamiltonian}. Dealing with two distinct vibrational bands, corresponding to upper and lower Dirac cones, the phonon Green function is a 2 × 2 square matrix.
For a better tractability, we select a constant attractive interaction independent of the phononic branch $g_\lambda=g$. 
Since the gap equation is written in units of $\hbar=k_B=m_e=1$, frequencies, momenta and temperature have energy dimensions, and we can replace them with dimensionless quantities normalized by the attractive interaction, i.e. $\overline{\omega}=\omega/g$.
The previous gap equation thus reduces to
\begin{equation}\label{eq:gap_eq}
    1=-\frac{4}{S\overline{\beta}} \sum_{\mathbf{k},m}
    \frac{i\overline{\nu}_m + \overline{\omega}_0}
    {\left\lbrack
    \overline{f}_{v_D}^2 + \left( \overline{\nu}_m-i\overline{\omega}_0 \right)^2
    \right\rbrack
    \left(
    \overline{\omega}_m^2+\overline{\xi}_\mathbf{k}^2+\overline{\Delta}^2
    \right)
    } \,,
\end{equation}
where $f_{v_D}=v_D \sqrt{k_x^2+k_y^2}$ stands for the Dirac cone-shaped dispersion relation that characterizes topological phonons. 
The Matsubara sum can be performed exactly, while the momentum sum can be replaced by the bi-dimensional integral given by periodic boundary conditions over the infinitely-extended topological surfaces.

\paragraph*{Numerical Results}

\begin{figure}[t]
    \centering
    \includegraphics[width=\linewidth]{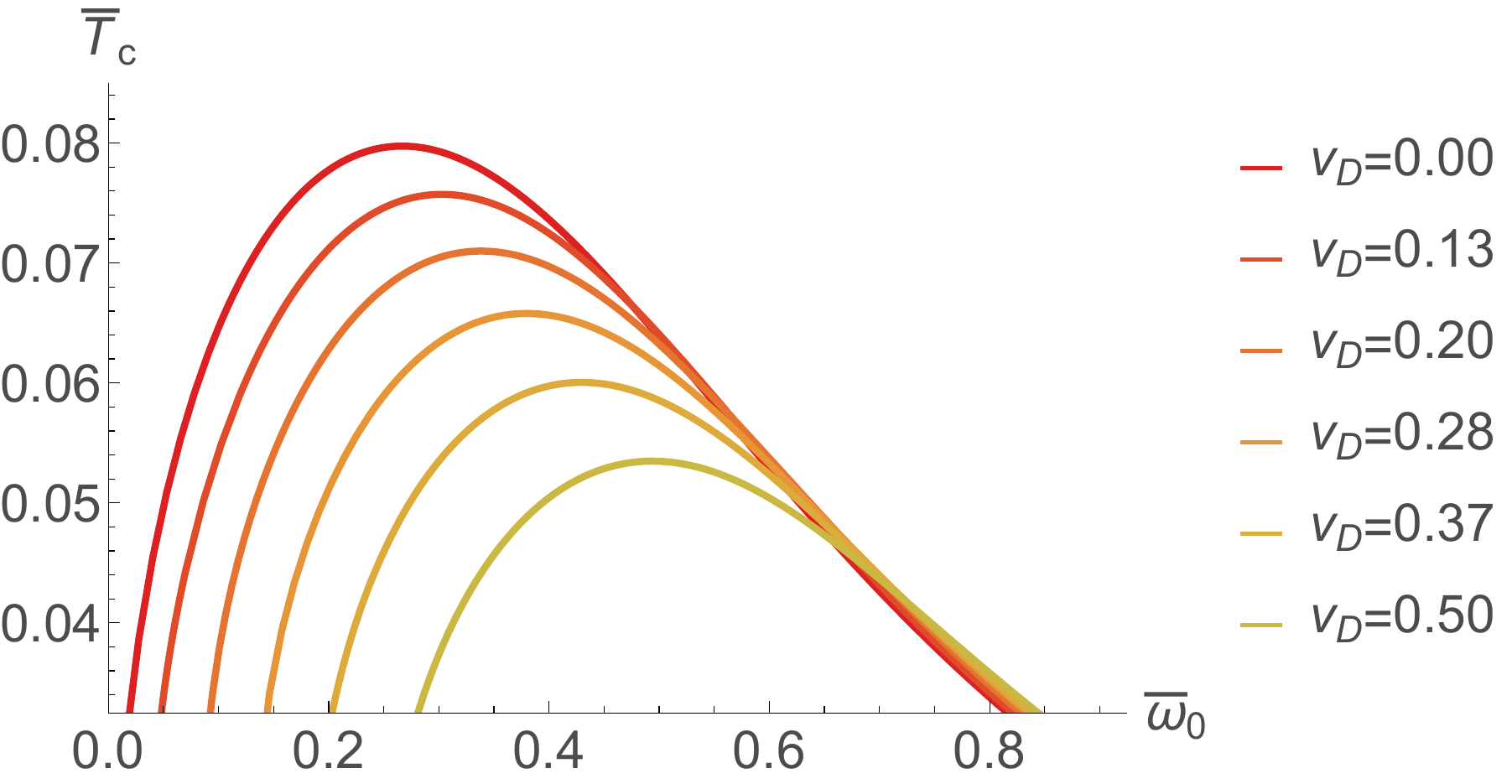}
    \caption{The dimensionless critical temperature $\overline{T}_c$ is displayed as a function of the frequency parameter $\overline{\omega}_0$, namely the phonon frequency at the Kramers-like point, for fixed chemical potential $\overline{\mu}=0.235$ and various Dirac velocities. For small $\overline{\omega}_0$, the structural instability related to negative phononic frequencies prevents finding solutions for $\overline{T}_c$.
    }
    \label{fig:temperature}
\end{figure}

A numerical solution for the dimensionless critical temperature $\overline{T}_{c}$ can be obtained by solving the above gap equation with a zero gap function $\overline{\Delta}=0$, and is plotted in Fig.\ref{fig:temperature} as a function of the dimensionless phononic frequency $\overline{\omega}_0$ at the Kramers-like point. 
For small frequency shifts, phonon states feature negative frequencies, therefore a lattice instability prevents finding solutions to the gap equation in that regime. This reflects on the summed function in Eq.\eqref{eq:gap_eq}, which has odd symmetry in the limit $\overline{\omega}_0=0$ and hence gives a zero contribution to the Matsubara sum.
The plot shows that $\overline{T}_c$ has a non-monotonic behaviour, reaching a maximum at an optimal frequency whose value depends on the slope $v_D$ of the Dirac cone-shaped topological bands. Specifically, the height of the peak decreases with increasing the band slope, thus the highest critical temperature is observed in correspondence to the limiting case $v_D=0$ where the Dirac cones degenerate into flat bands. 

\begin{figure}[b]
    \centering
    \includegraphics[width=\linewidth]{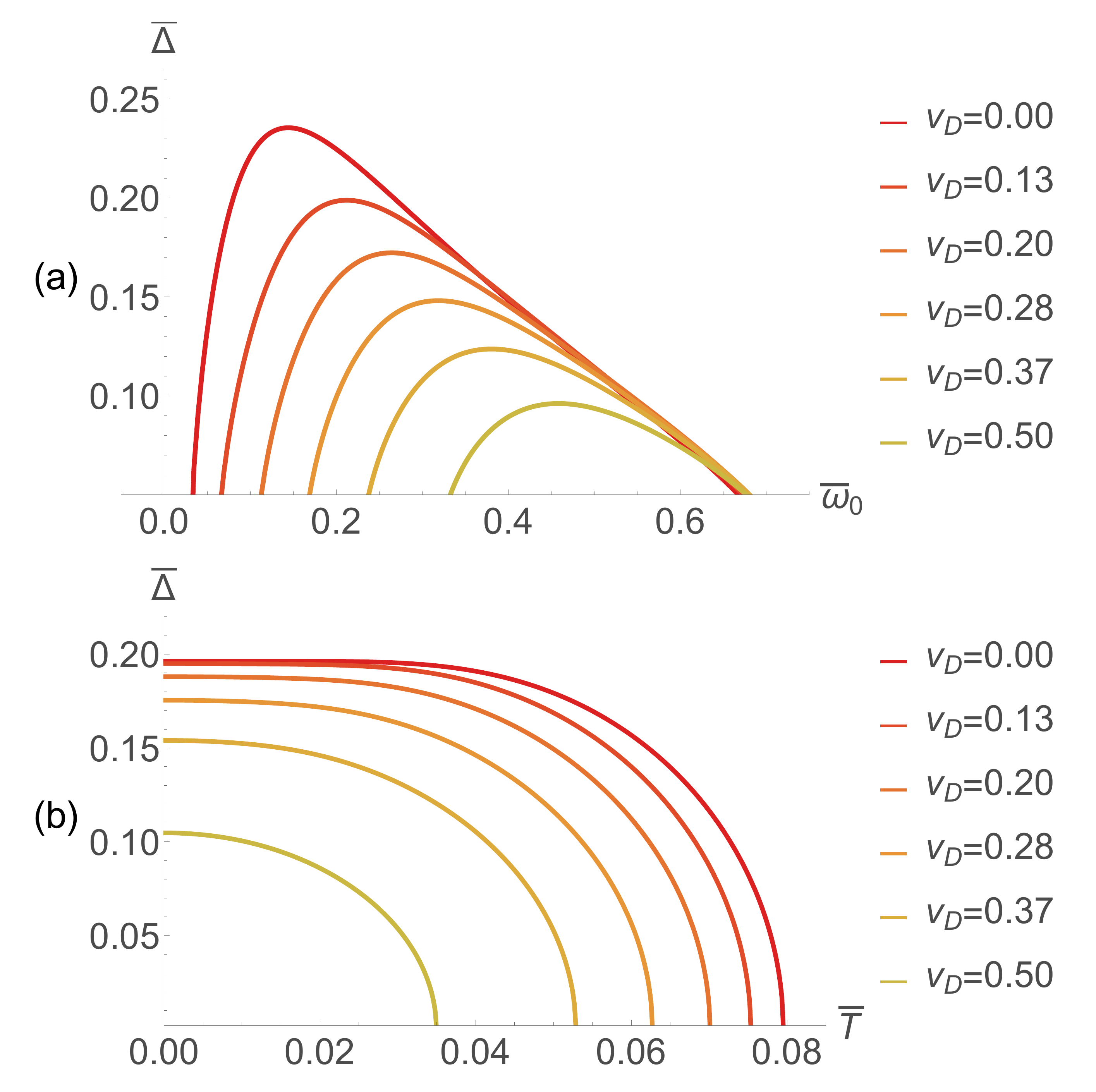}
    \caption{
    The dimensionless gap function is displayed (a) versus the frequency shift $\overline{\omega}_0$ for fixed temperature $\overline{T}=0.05$, and (b) as a function of temperature below the critical point for a fixed phonon frequency $\overline{\omega}_0=0.29$.
    In both plots, we set $\overline{\mu}=0.235$.
    For small $\overline{\omega}_0$, the structural instability related to negative phononic frequencies prevents finding solutions to the gap equation.
    }
    \label{fig:gap_function}
\end{figure}

The same behaviour is also shown by the dimensionless gap function, which represents the Cooper pairs binding energy. Numerical solutions, found by solving the gap equation for a fixed temperature below the critical value $\overline{T}_c$, are plotted in Fig.\ref{fig:gap_function} as a function of $\overline{\omega}_0$. 
Similarly to what happens with the critical temperature, an optimal value of the frequency at the Kramers-like point $\overline{\omega}_0$ determines an increase in the gap function, meaning that the bond between electrons in Cooper pairs is strengthened.
In Fig.\ref{fig:gap_function}, we also plot the gap function versus the dimensionless temperature, for a fixed value of $\overline{\omega}_0$. This figure shows the typical behaviour of a superconducting gap function, with the highest gap at zero temperature slowing down to zero at the critical point where the phase transition occurs.

\paragraph*{Physical Interpretation}

\begin{figure*}[t]
    \centering
    \includegraphics[width=\linewidth]{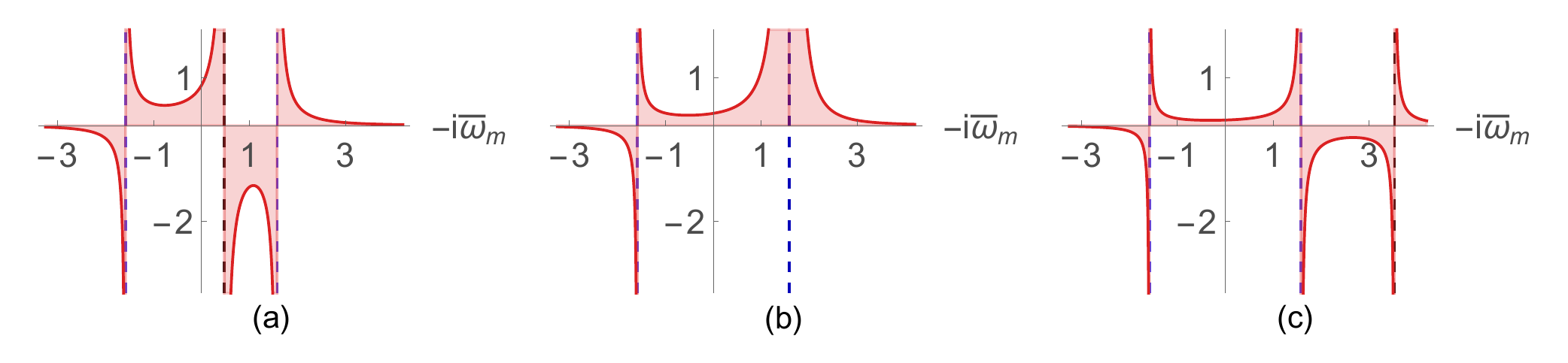}
    \caption{
    The summed function in Eq.\eqref{eq:gap_eq} is displayed as a function of the fermionic frequency $i\overline{\omega}_m$ for flat degenerate phononic bands with $v_D=0$ and fixed momentum within the Brillouin zone. The dimensionless frequency shift increases from left to right being (a) $\overline{\omega}_0<\overline{\xi}_\mathbf{k}$, (b) $\overline{\omega}_0=\overline{\xi}_\mathbf{k}$ and (c) $\overline{\omega}_0>\overline{\xi}_\mathbf{k}$.
    Black and blue dashed vertical lines represent different poles corresponding to electronic and phononic frequencies, respectively. 
    When $\overline{\omega}_0=\overline{\xi}_\mathbf{k}$, the \emph{overlap} between distinct poles maximizes the area covered by the function, meaning that there is a high contribution to the gap equation.
    }
    \label{fig:resonance}
\end{figure*}

The physical reason for the non-monotonic behaviour of $\overline{T}_c$ can be understood by studying the poles of the summed function, which give a high contribution to the Matsubara sum.
Taking the denominator into account and imposing $\overline{\Delta}=0$, the poles of the gap equation versus the Matsubara frequencies can be easily computed as
\begin{equation}
    -i\overline{\omega}_m = \pm \overline{\xi}_\mathbf{k} \,, \quad     
    -i\overline{\nu}_m = \overline{\omega}_0 \pm \overline{f}_{v_D} \,,  
\end{equation}
corresponding to the electronic and phononic energies, respectively. 
Varying $\overline{\omega}_0$, poles corresponding to the energy of distinct (quasi)particles may overlap, leading to an increased contribution to the Matsubara sum that reflects an enhanced superconducting coupling. At the overlap, phonons and electrons have the \emph{same} energy, meaning that the underlying reason for the superconducting enhancement is a \emph{resonance} between (quasi)particles.

The effect of such a resonance can be clearly seen in the instructive case of degenerate phonons. The summed function for flat bands with $v_D=0$ is displayed in Fig.\ref{fig:resonance} versus the Matsubara fermionic frequency $i\overline{\omega}_m$. The Kramers-like point frequency $\overline{\omega}_0$ increases from left to right.
Blue and black dashed lines represent the poles of the function, corresponding to electronic $\overline{\xi}_\mathbf{k}$ and phononic $\overline{\omega}_0 \pm \overline{f}_{v_D}$ energies, respectively. 
Due to the energy degeneration $\overline{f}_{v_D}=0$, and hence there is only a single pole corresponding to both phonon states.
For fixed momentum, the resonance condition is given by $\overline{\omega}_0=\overline{\xi}_\mathbf{k}$, which ensures that electrons and phonons over the topological surfaces have the same energy.
Herein, electronic and phononic poles overlap, and the part of the plot with negative area included between them vanishes. Therefore, the overlap of the poles maximizes the positive area covered by the summed function, thus giving a large contribution to the gap equation.
For $\overline{\omega}_0$ values away from this overlap, there is a wide range of frequencies that contributes negatively to the gap equation. 
This means that, for selected momentum inside the Brillouin zone boundary, there is a non-monotonic behaviour peaked at the value of $\overline{\omega}_0$ that guarantees a resonance between electrons and phonons. The overall effect on the critical temperature is then given by a sort of ``momentum average'' produced by the momentum integral over the Brillouin zone boundary.

A very similar behaviour is featured also by non-flat phonons (i.e. with a finite slope of the Dirac cone), whose energy states are split by a non-zero Dirac velocity. 
Due to the energy splitting, there are two distinct overlaps, given by $\overline{\omega}_0 = \overline{\xi}_\mathbf{k} \pm \overline{f}_{v_D}$ for fixed momentum, where electrons and phonons have the same energy.
Herein, the area covered by the summed function is maximized, but the existence of a secondary phononic pole, not involved in the resonance with electron states, implies the presence of a range of Matsubara frequencies with a negative contribution to the gap equation.
Increasing the slope, and hence the splitting between phonons at the same momentum state, widens such a negative range. 
This means that the overall effect of resonance decreases with increasing the slope $v_D$ of the Dirac cone-shaped states, in agreement with the numerical results shown in Figs.\ref{fig:temperature}-\ref{fig:gap_function}.
Away from the resonances, the gap equation behaves in the same way as in the previous degenerate case, with a wide range of Matsubara frequencies, included between electronic and phononic poles, that give a negative contribution to the gap equation and this, in turn, decreases the electron pairing.

So far, several materials with topological phononics physics have been reported \cite{TP_Oxide_Perovskites, TP_Weyl}, but none 3D topological insulating-like state have been experimentally realized.
A candidate material should be a crystal with lattice symmetries able to produce degenerate Kramers-like states along the high-symmetry lines \cite{crystalline_TIs}. Such symmetries should also be preserved when projected onto the surfaces where edge states arise.
Crystals with $C_{nv}$ symmetries for $n=3,4,6$ show interesting physics for electrons \cite{Spin-Orbit_free_TIs}, and could be similarly exploited to realize phononics topological insulators. 
Also different kinds of symmetries, for instance magnetic space group symmetry, particle hole-symmetry and their combinations, could be used to define novel pseudospin degrees of freedom \cite{phononic_TIs}, in order to control phonons and construct topological insulating-like phases.
A 2D superconductor, like the ultrathin lead film grown on a semiconductor substrate \cite{2D_Superconductor_Pb-Si, 2D_Superconductor_Pb,2D_superconductor_Pb-GaAs}, can then be placed on top of their topological surfaces: the interface between them  contains the proposed physics.

To conclude, we have developed a theory of superconductivity where the superconducting states are mediated by boundary Dirac-like phonons at the interfaces of topological phononics insulators.
The surface phonon dynamics is accounted for by a suitable propagator implemented into a self-consistent two-band gap equation for the Cooper pairing between electrons. 
Upon solving the gap equation, it is found that the $T_c$ depends non-monotonically on the phononic frequency $\omega_0$ at the Kramers-like point and features a maximum as a function of $\omega_0$. 
A strong dependence on the slope of the Dirac cone-shaped bands is also observed, with the highest peak of critical temperature that occurs in correspondence of degenerate flat-band topological phonons. 
The value of the frequency parameter around which Cooper pairing is the strongest is set by the effect of a resonance between the (standard) electrons and the topological phonons. Within this optimal range of frequencies, the strongly enhanced electron-phonon interaction increases the superconducting coupling between electrons.
Outside this window, instead, the strength of pairing deteriorates, leading to a reduction in $T_c$. 
In addition, the electron-phonon resonance is greatly enhanced in the limit of flat degenerate phononic bands, because both phonon states are then simultaneously involved in the interaction with electrons. Conversely, a non-zero slope of the Dirac cone splits the energy states, thus decreasing the overall superconducting pairing.
It is hoped that this work will encourage the development of a new route for enhancing the superconducting $T_c$ and stimulate the search for phononic topological insulators in real materials.


\bibliographystyle{apsrev4-1}
\bibliography{reale}

\end{document}